\documentclass{aa}
\usepackage{graphicx}
\usepackage{lscape}
\usepackage{txfonts}
\usepackage{color}

\begin{document}


\title{The Araucaria Project. Precise physical parameters of the eclipsing
binary IO~Aqr}
\titlerunning{The eclipsing binary IO~Aqr}
\author{D.~Graczyk\inst{1,2},
P. F. L.~Maxted\inst{3},
G.~Pietrzy\'nski\inst{4,2},
B.~Pilecki\inst{4,2},
P.~Konorski\inst{4},\\
W.~Gieren\inst{2,1},
J.~Storm\inst{5},
A.~Gallenne\inst{2},
R. I.~Anderson\inst{6,7},
K.~Suchomska\inst{4},\\
R. G.~West\inst{8},
D.~Pollacco\inst{8},
F.~Faedi\inst{8}
\and G.~Pojma\'nski\inst{4}}
\authorrunning{D. Graczyk et al.}
\institute{Millenium Institute of Astrophysics, Santiago, Chile 
\and Universidad de Concepci\'on, Departamento de Astronom\'ia, Casilla 160-C, Concepci\'on, Chile
\and Astrophysics Group, Keele University, Staffordshire, ST5 5BG, UK
\and Warsaw University Observatory, Al. Ujazdowskie 4, 00-478 Warsaw, Poland
\and Leibniz-Institut f\"{u}r Astrophysik Potsdam, An der Sternwarte 16, 14482 Potsdam, Germany
\and Observatoire de Gen{\`e}ve, Universit{\'e} de Gen{\`e}ve, 51 Ch. des Maillettes, 1290 Sauverny, Switzerland
\and Department of Physics and Astronomy, Johns Hopkins University, Baltimore, MD 21218, USA
\and Department of Physics, University of Warwick, Coventry CV4 7AL, UK
}




\abstract
{}
{Our aim is to precisely measure the physical parameters of the eclipsing binary IO~Aqr and derive a distance
to this system by applying a surface brightness - colour relation. Our motivation is 
to combine these parameters with future precise distance determinations from the GAIA space mission
to derive precise surface brightness -- colour relations for stars.}
{We extensively used photometry from the Super-WASP and ASAS projects and precise
radial velocities obtained from HARPS and CORALIE high-resolution spectra. We
analysed light curves with the code JKTEBOP and radial velocity curves with the Wilson-Devinney  program.}
{We found that IO~Aqr is a hierarchical triple system consisting of a double-lined short-period
($P=2.37$ d) spectroscopic binary and a low-luminosity and low-mass companion 
star orbiting the binary with a period of $\gtrsim 25000$ d ($ \gtrsim70$ yr) on a very eccentric orbit. 
We derive high-precision (better than $ 1\%$) physical parameters of the inner 
binary, which is composed of two slightly evolved main-sequence stars (F5 V-IV
+ F6 V-IV) with masses of $M_1 = 1.569\pm0.004$ and $M_2=1.655\pm0.004$~M$_\odot$
and radii $R_1 = 2.19\pm0.02$ and $R_2=2.49\pm0.02$~R$_\odot$. The companion
is most probably a late K-type dwarf with mass $\approx0.6$~M$_\odot$. The
distance to the system resulting from applying a ($V\!-\!K$) surface
brightness -- colour relation is 255 $\pm$ 6 (stat.) $\pm$ 6 (sys.) pc, which agrees well with the Hipparcos value of 
$270^{+91}_{-55}$ pc, but is more precise by a factor of eight.}
{}


\maketitle

\section{Introduction}
Detached eclipsing binary stars are a very important source of precise stellar parameters, especially of absolute radii and masses.
It has been proven that such parameters combined with empirical stellar surface brightness - colour (SBC)
relations derived from interferometric measurements \citep[e.g.][]{ker04,cha14} result in accurate distance determinations even 
to extragalactic eclipsing binaries. Indeed, very accurate distances to the two Magellanic Clouds determined with the eclipsing binary method have recently been reported \citep{pie13,gra14}. In this method, the SBC relation is used as a tool for predicting the angular diameters of the system components \citep{lac77}. The distance is simply derived by comparing the absolute diameter of a component with its predicted angular diameter. The method involves simple geometric considerations and is therefore a powerful tool for measuring distances with a minimum of modelling assumptions. 

However, the procedure may be reversed: using prior knowledge of the distance to a particular detached eclipsing binary, 
SBC empirical relations can be derived in a way completely independent of interferometry. This basic idea has long been known and was firstly used by \cite{ste10,ste11} to derive the surface brightness of the two components of Algol and $\beta$ Aur. 
Within about 250 pc of the Sun there are $\sim\!100$ confirmed detached eclipsing binaries with parallaxes measured by the Hipparcos mission \citep[e.g.][]{kru99,tor10}. However, about half of them still lack a detailed analysis of their physical parameters or any analysis at all. These systems comprise components covering a wide range of spectral types from B1 to K0 that is suitable for calibrating precise SBC relations in the corresponding colour ranges. The are several factors that influence the precision of SBC calibrations, most importantly, 1) the uncertainty on absolute dimensions, especially the radii, 2) trigonometric parallax errors, 3) intrinsic variability of the star, and 4) zero-point shifts between different photometric systems. 

Recently, a few catalogues of precise absolute dimension determinations from eclipsing binaries have been published \citep{tor10,eke14,sou14}
that cover a wide range of temperatures. However, these catalogues show that for the purpose of SBC calibration, a precise analysis of new eclipsing binaries is needed.  For example, \citet[][see their Fig. 13]{tor10} listed 15 components from nine
systems (EI Cep, V422 Cyg, RZ Cha, GX Gem, BW Aqr, DM Vir, CD
Tau, AI Phe, and V432 Aur) in a region of $T_{\rm eff}$ -- $\log{g}$ plane covering a box of 6000 -- 7000 K, 3.6 -- 4.1 dex (F-type subgiants). Despite the number of systems with well-determined masses and radii in this region of parameter space, the calibration of the SBC for these stars is not very reliable because one system lacks a trigonometric parallax (V442 Cyg), three systems have very low quality trigonometric parallaxes (GX Gem, BW Aqr, and DM Vir), the Hipparcos parallax of another system (AI Phe) is inconsistent with the value derived using the eclipsing binary method \citep{and88}, and one system is magnetically active with distortions to the light curve from star spots (V432 Aur). This leaves only five components in three non-active, detached eclipsing binaries with precise absolute dimension determinations within 250 pc from the Sun, and with secure trigonometric parallaxes.   

It is believed that parallax uncertainties from Hipparcos are dominated by statistical errors  -- photon statistics \citep[e.g.][]{vLe07b}. For F-type eclipsing binary with a parallax of 4 mas we expect an error of about 1 mas, or in other words, a relative precision of 25~\%. To improve this, we could average results from many components over colour bins because the statistical uncertainty would decrease as $\sqrt{N}$, where $N$ is the number of components used in analysis.

The Gaia mission \citep{per01} is expected to much improve the Hipparcos distances by minimising statistical uncertainties, and for the F-type system mentioned before, the uncertainty of the parallax should be at most 0.5~\% \citep[][]{deB15}. The SBC calibration would benefit from such precise and accurate Gaia parallaxes obtained for a larger number of detached eclipsing binaries. The photometric distances derived for many of them would also be used as an independent check of GAIA results (if some significant systematic were to occur).

Inhomogeneous photometry causes difficulties with transformations between different photometric systems and zero-point shift uncertainties. The situation somewhat resembles problems occurring with the
accurate determination of temperatures and angular diameters using the infrared flux method, where the main error comes from photometric calibration uncertainties \citep[see discussion in][]{cas14}. Anticipating highly accurate parallaxes from Gaia (precision better than 1\%) and the standard precision of present-day radii determination (better then 2\%), the systematic error of future SBC calibration would be dominated by photometric and radiative properties of stellar atmosphere uncertainties (Andersen, priv. communication).

Here, we present a detailed analysis of IO~Aqr (HD~196991,
HIP~102041, $\alpha_{2000}\!=\!20^h 40^m 45^s\!\!.5$, $\delta_{2000}\!=\!00^{\circ} 56^{'} 21^{''}$), 
an eclipsing binary system composed of two mid-F-type stars located in 
the middle of the $T_{\rm eff}$ -- $\log{g}$ plane of the region
described above. Despite
being relatively bright  ($V\sim8.8$~mag), this star was not recognised as an
eclipsing binary until it was observed with Hipparcos and assigned its variable
star name by \cite{kaz99}. Time-series photometry is also available from the
Wide-Angle Search for Planets \citep[WASP;][]{pol06}. During our analysis we
discovered that IO~Aqr is a triple system, where a close inner
binary has a low-mass companion star on a much wider orbit. The system was
analysed before by \cite{dim04}. They derived a set of fundamental parameters
for the binary, but did not detect any trace of the tertiary component. Although the companion complicates the interpretation of observations for
IO~Aqr, the companion is optically faint (less than 1\% of the total light in
V) and, being on a relatively wide orbit, exerts only long-term perturbations on
the inner system, so we have been able to determine precise parameters for
both stars in the eclipsing binary component of this hierarchical triple
system.

\section{Observations}

\subsection{Photometry}
\subsubsection{ASAS}
The $V$-band photometry of IO~Aqr, publicly available from the ASAS
catalogue\footnote{\texttt{http://www.astrouw.edu.pl/asas/?page=aasc}}
\citep[ACVS;][]{poj02}, spans from 2001 March 27 to 2009 December 1 and
contains 384 good-quality points (flagged ``A'' in the original data). This
photometry was extended by 266 measurements from ASAS-North obtained 
in a second ASAS station on Haleakala, Hawaii.  We also have access to
previously unpublished $I$-band photometry from the ASAS and ASAS-North
catalogues that spans from 1998 August 21 to 2009 June 4 and contains 1174
good points. In total we have four ASAS data sets, two in V band and two in I band.

\subsubsection{WASP}
 IO~Aqr (1SWASP~J204045.47+005621.0) is one of several million bright stars
($8\la V \la 13$)  that have been observed by WASP. The WASP survey is described in \citet{pol06}
and \citet{2008ApJ...675L.113W}. The survey obtains images of the night sky
using two arrays of eight cameras, each equipped with a charge-coupled device
(CCD), 200-mm f/1.8 lenses and a broad-band filter (400\,--\,700\,nm).  Two
observations of each target field are obtained every 5\,--\,10 minutes using
two 30\,s exposures. The data from this survey are automatically processed and
analysed to identify stars with light curves that contain transit-like
features that may indicate the presence of a planetary companion. Light curves
are generated using synthetic aperture photometry with an aperture radius of
48\,arcsec. The data for all the stars in each target field are processed
using the {\sc sysrem} algorithm \citep{2005MNRAS.356.1466T} to produce
differential magnitudes that are partly corrected for systematic noise trends in the
photometry. Data for this study were obtained between 2006 July 4 and 2010
October 6.

\subsubsection{HIPPARCOS}
We downloaded the Hipparcos photometry in $H_P$
magnitudes from a public archive\footnote{\texttt{https://www.rssd.esa.int/index.php?project=HIPPARCOS
\&page=Epoch\_Photometry}} . We removed four discrepant high points from the data set and finally had 87
photometric points in total.  

\subsection{Spectroscopy}
\subsubsection{HARPS}
We obtained spectra of IO~Aqr  with the High Accuracy Radial velocity
Planet Searcher \citep[HARPS;][]{may03} on the European
Southern Observatory 3.6 m telescope in La Silla, Chile. IO~Aqr is a bright
target, therefore observations were generally obtained in marginal observing
conditions or during twilight. Observations were obtained between 2009 October
10 and 2014 September 10. A total of 17 spectra were secured in high-efficiency (``EGGS'') mode. The exposure times were typically 260\,s, resulting
in an average signal-to-noise ratio (S/N) per pixel of $\sim60$. All spectra were
reduced on-site using the HARPS data reduction software (DRS).

\subsubsection{CORALIE}
Twenty spectra were obtained with the CORALIE spectrograph on the Swiss 1.2 m
Euler telescope at La Silla observatory, Chile, between 2008 October
5 and 2010 November 1. The exposure times varied from 780\,s to 1200\,s,
giving a typical S/N near 5500\AA\ of 50 per pixel. The spectra were reduced 
using the automated reduction pipeline. 

\section{Analysis}

\subsection{JKTEBOP \label{sect_jktebop}}

 We used \textsc{jktebop}\footnote{\texttt
{www.astro.keele.ac.uk/$\sim$jkt/codes/jktebop.html}} version 25
(\citealt{sou04a,sou04b} and references therein) to perform least-squares fits
of the {\sc ebop} light-curve model \citep{pop81} to the various photometric
data sets.

 For the epoch of primary eclipse for each photometric data set, $T_0$, we used
the time of mid-eclipse that was closest to the average time of observation for that
data set. Other free parameters in the least-squares fit were a normalisation
constant, the surface brightness ratio $J = S_2/S_1$, where $S_1$ and $S_2$ are the
surface brightness of the primary and the secondary star, respectively; 
the sum of the radii relative to semi-major axis, $r_1+r_2
=(R_1+R_2)/a$; the ratio of the radii, $k=r_2/r_1$; the orbital inclination,
$i$ and the orbital period, $P_{\rm orb}$. The mass ratio was fixed at the
value derived from the spectroscopy described below. We consulted various
tabulations of linear limb-darkening coefficients
\citep{VanHam93,DiazCo95,Claret95,Claret00} and then set the value of $x_{\rm
LD}$ for both stars to be equal to a fixed average value from these
tabulations. We performed independent least-squares fits to investigate the
sensitivity of the other parameters to the assumed value of $x_{\rm LD}$ and
accounted for this additional uncertainty when calculating the standard errors
on these parameters. We assumed that the orbit is circular and fixed the
gravity-darkening coefficients of both stars to 0.2 -- this parameter has a
negligible effect on the results.

The luminosity ratio and third light contribution estimated from the analysis
of the spectroscopic data were included as constraints in the least-squares
fits. The relative weightings of the data were set equal to the root-mean-square
(rms) residual of a preliminary fit to each data set. We used the residual
permutation method to estimate robust standard errors on the free parameters.
The results are given in Table~\ref{tab_lc_jkt}, and the fits to the individual
light curves are shown in Fig.~\ref{fig_lcfit}. Further details of these fits
are given below.

\subsubsection{WASP photometry \label{sect_wasp_lcfit}}
 We divided the WASP data into eight subsets, each subset comprising the data from
one camera and one observing season.  Outliers were identified using the
residuals from a preliminary light curve model and were removed from the analysis.
The number of data points used in the analysis and observation dates of each
subset are given in Table~\ref{tab_wasp_subsets}. We identified between 6 and
25 nights in each subset of data where there was an offset between the data
observed on that night and the rest of the data. We modified our version of
\textsc{jktebop}  to include an offset for the data obtained on each of these nights as
additional free parameters in the least-squares fit.  The values and standard
errors given in Table~\ref{tab_lc_jkt} are the weighted mean and standard
error of the mean for the results from these eight least-squares fits. The times
of mid-eclipse derived from these least-squares fits are given in
Table~\ref{tab_tmin}.

\subsubsection{ASAS photometry}
 We fitted the four ASAS data sets independently. These data span a wide range
of dates of observation with only one observation per night. The effect of the
period variations discussed below on the parameters derived is negligible.
To derive times of mid-primary eclipse for the period analysis, we divided the
data into subsets, each covering approximately 200 nights. We then performed
least-squares fits with only $T_0$, $J,$ and a scaling factor as free
parameters to derive the  times of mid-primary eclipse given in
Table~\ref{tab_tmin}.

\subsubsection{Hipparcos photometry}
 We used Hipparcos photometry to derive the single time of
mid-primary eclipse given in Table~\ref{tab_tmin} using the same method as we
used for the ASAS data. 

\begin{figure}
\mbox{\includegraphics[width=0.47\textwidth]{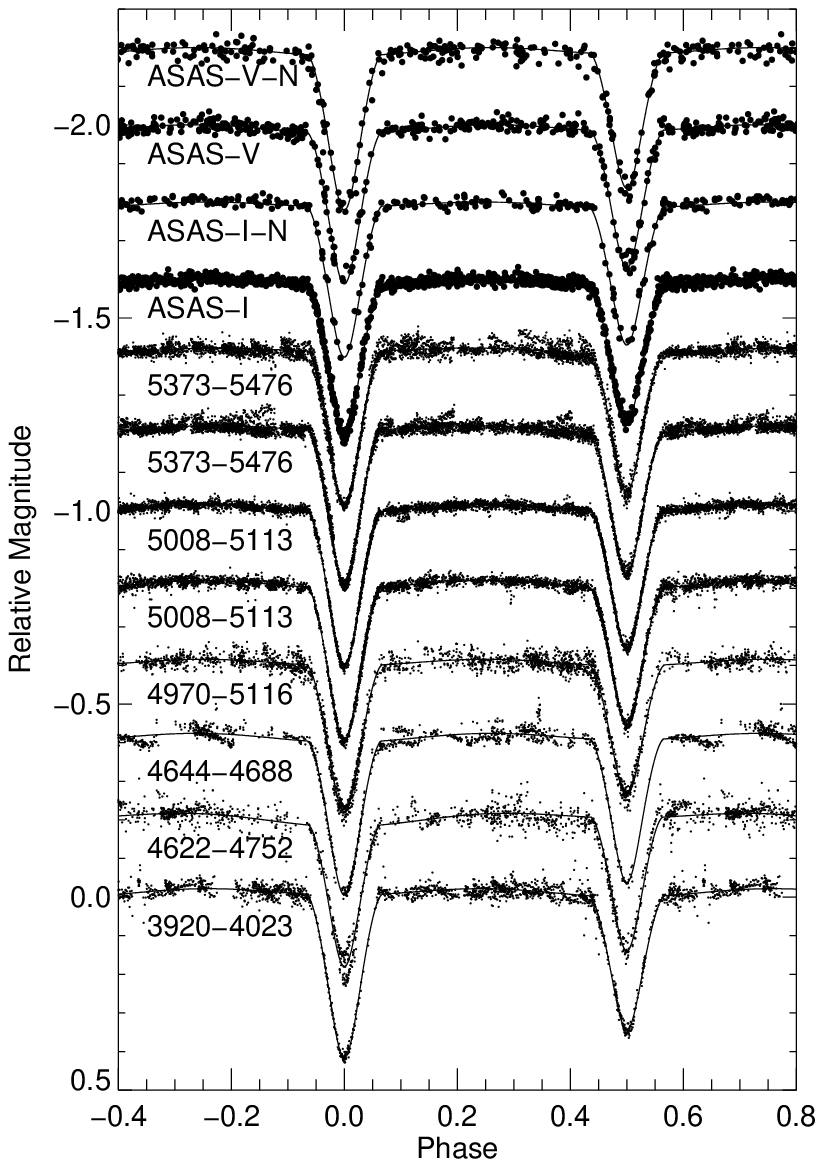}}
\caption{From bottom to top: WASP,  ASAS I-band, and ASAS V-band photometry of
IO~Aqr with \textsc{jktebop} light-curve model fits. The numbers are the ranges of Julian Dates for WASP data. 
\label{fig_lcfit}}
\end{figure}

\begin{table}
\centering
\caption{Log of WASP photometry for IO~Aqr. The start and end dates of observation, $t_{\rm
start}$ and $t_{\rm end}$, are JD-2450000.  $N_{\rm ok}$ is the number of data
points used in this analysis, and $N_{\rm block}$ is the number of
nights of data that were fitted with an independent zero point in the
least-squares fit. }
\label{tab_wasp_subsets}
\begin{tabular}{@{}rrrr@{}}
\hline \hline
$ t_{\rm start}$ & 
$t_{\rm end}$ & 
 $N_{\rm ok}$ & $N_{\rm block}$ \\
\hline
3920 & 4024 & 1865  &  17\\
4622 & 4753 & 1339  &   9\\
4644 & 4689 & 1094  &   6\\
4970 & 5117 & 2251  &   6\\
5008 & 5114 & 3825  &  25\\
5008 & 5114 & 4944  &  18\\
5373 & 5477 & 5699  &  22\\
5373 & 5477 & 4522  &  25\\
\hline
\end{tabular}
\end{table}

\begin{figure}
\mbox{\includegraphics[width=0.47\textwidth]{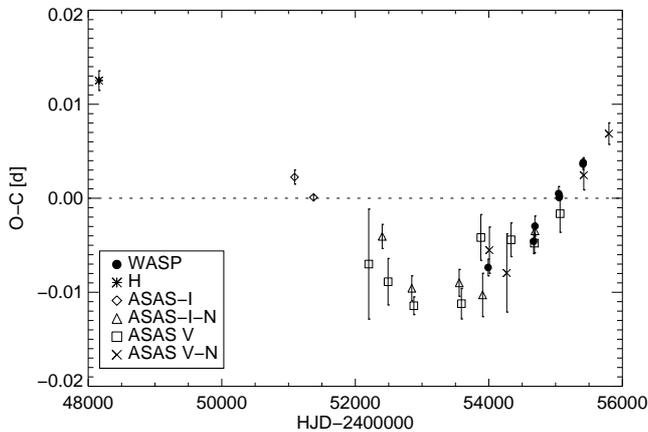}}
\caption{Residuals from the best-fit linear ephemeris to our measured times
of mid-primary eclipse.
\label{fig_ephem}}
\end{figure}

\begin{table}
\centering
\caption{Times of mid-primary eclipse for IO~Aqr  expressed in Heliocentric Julian Date.}
\label{tab_tmin}
\begin{tabular}{@{}rrl@{}}
\hline \hline
Epoch &  HJD  & Source \\
\hline
-2485 & 2448161.3007$\pm$   0.0010  & Hipparcos \\
-1248 & 2451090.6515$\pm$   0.0008  & ASAS-I \\
-1128 & 2451374.8234$\pm$   0.0002  & ASAS-I \\
 -694 & 2452402.5822$\pm$   0.0013  & ASAS-I-N \\
 -507 & 2452845.4146$\pm$   0.0013  & ASAS-I-N \\
 -207 & 2453555.8503$\pm$   0.0014  & ASAS-I-N \\
  -59 & 2453906.3304$\pm$   0.0023  & ASAS-I-N \\
  272 & 2454690.1840$\pm$   0.0008  & ASAS-I-N \\
 -779 & 2452201.2893$\pm$   0.0059  & ASAS-V \\
 -657 & 2452490.1977$\pm$   0.0025  & ASAS-V \\
 -494 & 2452876.1983$\pm$   0.0009  & ASAS-V \\
 -193 & 2453589.0018$\pm$   0.0016  & ASAS-V \\
  -70 & 2453880.2872$\pm$   0.0024  & ASAS-V \\
  121 & 2454332.5974$\pm$   0.0018  & ASAS-V \\
  269 & 2454683.0784$\pm$   0.0010  & ASAS-V \\
  431 & 2455066.7165$\pm$   0.0020  & ASAS-V \\
  -16 & 2454008.1642$\pm$   0.0025  & ASAS-V-N \\
   94 & 2454268.6547$\pm$   0.0042  & ASAS-V-N \\
  581 & 2455421.9382$\pm$   0.0016  & ASAS-V-N \\
  740 & 2455798.4733$\pm$   0.0011  & ASAS-V-N \\
  903 & 2456184.4783$\pm$   0.0017  & ASAS-V-N \\
  -24 & 2453989.2174$\pm$   0.0009  & WASP\\
  264 & 2454671.2380$\pm$   0.0013  & WASP\\
  272 & 2454690.1845$\pm$   0.0011  & WASP\\
  422 & 2455045.4056$\pm$   0.0007  & WASP\\
  425 & 2455052.5095$\pm$   0.0003  & WASP\\
  427 & 2455057.2459$\pm$   0.0002  & WASP\\
  576 & 2455410.0988$\pm$   0.0006  & WASP\\
  577 & 2455412.4671$\pm$   0.0005  & WASP\\
  \hline
\end{tabular}
\end{table}

\begin{table*}
\centering
\caption{Results of the \textsc{jktebop} fit to the observed LCs. Values for the WASP
data  are the weighted means and standard errors from 8 subsets, as described
in Sect.~\ref{sect_wasp_lcfit}. Symbols are defined in Sect.~\ref{sect_jktebop}.
Values preceded by "=" are fixed parameters in the least-squares fit. Values
in parentheses are standard errors on the final digits of the quantity shown.
Values in square brackets were constrained within the limits indicated during
the least-squares fit. Where errors are given on fixed parameters, 
independent least-squares fits were performed to calculate the contribution 
of the uncertainty shown to the standard errors quoted on the other parameters.}
\label{tab_lc_jkt}
\begin{tabular}{lrrrrrr}
\hline \hline
Parameter&
\multicolumn{1}{l}{WASP}&
\multicolumn{1}{l}{ASAS-I}&
\multicolumn{1}{l}{ASAS-I-N}&
\multicolumn{1}{l}{ASAS-V}&
\multicolumn{1}{l}{ASAS-V-N}&
\multicolumn{1}{l}{Mean}\\
\hline
$T_0$ ( HJD-2450000) & ---          & 1396.13639(24) & 3589.0060(6) & 3589.0052(6) & 5315.3727(21) &        --- \\
$P_{\rm orb} $ d       & 2.368145(17) &  2.3681012(51) &  2.368121(2) &  2.368123(2) &   2.368136(3) &        --- \\
$r_1+r_2$         &   0.4238(13) &     0.4256(31) &   0.4277(56) &   0.4192(48) &    0.4243(63) & 0.4244(14) \\
$k = r_2/r_1$      &    1.1500(100) &     1.1462(39) &   1.1451(89) &   1.1551(78) &   1.1746(110) & 1.1478(36) \\
$i$ (deg)          &     81.44(8) &      81.47(18) &    81.43(26) &    81.74(22) &     81.45(28) &   81.47(7) \\
$r_1$              &    0.1984(9) &     0.1938(17) &   0.1988(11) &   0.1945(25) &    0.1951(33) &  0.1976(18) \\
$r_2$              &   0.2265(18) &     0.2273(15) &   0.2268(19) &   0.2247(24) &    0.2292(33) &  0.2268(16) \\
$J=S_2/S_1$        &     0.908(6) &       0.929(5) &     0.911(4) &    0.918(12) &     0.888(18) &        --- \\
$x_{\rm LD}$           &        =0.6  &       =0.42(9) &     =0.42(9) &     =0.60(8) &      =0.60(8) &        --- \\
$L_3$ (\%)         &     [1.0(5)] &       [1.0(5)] &     [1.0(5)] &     [1.0(5)] &      [1.0(5)] &        --- \\
$L_2/L_1$          &    [1.21(4)] &      [1.23(3)] &    [1.23(3)] &    [1.21(3)] &     [1.21(3)] &        --- \\
$rms$ (mag)     & 0.014&0.009&0.011&0.014&0.018& --- \\
\hline
\end{tabular}
\end{table*}

\subsection{Period variations \label{per_var}}
 We performed a least-squares fit to data given in Table~\ref{tab_tmin} to derive
the following linear ephemeris for the time of mid-primary eclipse in Heliocentric Julian Date:
\[ {\rm HJD~(UTC)} = 2\,454\,046.0596(6) +   2.368\,117\,2(8)\cdot E. \]
The standard error on the final digit of each term is given in parentheses.
The residuals from this ephemeris are shown in Fig.~\ref{fig_ephem}. It is clear that the period of this binary varies. Assuming that these changes are periodic, we can estimate their minimum period to be $> 8000$ d. This is discussed further in Sect.~\ref{wd}.

\subsection{Radial velocities \label{rad_vel}}

\begin{table}
\centering
\caption{RV measurements for IO~Aqr. Index ``1''
denotes the hotter star (primary), ``2'' is the cooler (secondary), and ``3" is the tertiary component. 
 Numbers in parenthesis give the uncertainty. BJD means the Barycentric Julian Date and CORAL means the CORALIE spectrograph. }
\label{tab_rv}
\begin{tabular}{@{}lrrrr@{}}
\hline \hline
BJD & $RV_1$ &  $RV_2$ & $RV_3$ & Spectr. \\
-2450000& (km s$^{-1}$) & (km s$^{-1}$)&(km s$^{-1}$) & \\
\hline
4745.49326 &   -84.34(20) &    99.64(23) & 3.0(1.1) & CORAL \\
4746.59946 &   117.50(20) &   -91.90(23) & 1.3(1.1) & CORAL \\
4746.70296 &    98.85(20) &   -74.51(23) & 4.6(1.1) & CORAL \\
4747.49989 &  -104.83(20) &   118.10(23) & 3.1(1.1) & CORAL \\
4747.63290 &  -110.13(20) &   123.30(23) & 3.0(1.1) & CORAL \\
5086.57856 &   -68.78(20) &    83.75(23) & 1.1(1.1) & CORAL \\
5086.59403 &   -64.90(20) &    80.74(23) & 2.0(1.1) & CORAL \\
5087.55728 &   123.67(20) &   -98.49(23) & 1.4(1.1) & CORAL \\
5087.66076 &   109.02(20) &   -85.24(23) & --- & CORAL \\
5088.54850 &  -107.94(20) &   120.58(23) & 1.8(1.1) & CORAL \\
5088.63650 &  -110.38(20) &   122.96(23) & 2.2(1.1) & CORAL \\
5089.55659 &   103.59(20) &   -79.53(23) & 2.3(1.1) & CORAL \\
5089.64296 &   117.78(20) &   -93.71(23) & 0.9(1.1) & CORAL \\
5090.55337 &   -37.84(20) &    54.57(23) & --- & CORAL \\
5090.62798 &   -58.25(20) &    74.19(23) & 0.3(1.1) & CORAL \\
5120.57639 &   129.38(15) &  -103.81(17) & 2.0(1.0) & HARPS \\
5447.66922 &    99.38(15) &   -75.39(17) & 2.4(1.0) & HARPS \\
5447.75236 &    79.77(15) &   -56.73(17) & 2.2(1.0) & HARPS \\
5449.50961 &   103.02(15) &   -79.17(17) & 1.3(1.0) & HARPS \\
5467.49889 &  -110.02(15) &   122.78(17) & 1.9(1.0) & HARPS \\
5468.50664 &   112.30(15) &   -88.17(17) & 1.5(1.0) & HARPS \\
5469.51563 &   -54.62(15) &    70.93(17) & 1.7(1.0) & HARPS \\
5479.51122 &  -101.29(15) &   115.13(17) & 2.0(1.0) & HARPS \\
5499.54423 &   128.36(20) &  -103.11(23) & 2.4(1.1) & CORAL \\
5499.57439 &   127.18(20) &  -101.40(23) & 3.6(1.1) & CORAL \\
5500.53761 &  -101.57(20) &   114.93(23) & 0.5(1.1) & CORAL \\
5500.57142 &  -105.01(20) &   118.33(23) & 1.3(1.0) & CORAL \\
5501.53207 &    85.99(20) &   -62.50(23) & --- & CORAL \\
5536.52398 &   -68.88(15) &    84.30(17) & 0.7(1.0) & HARPS \\
6572.48783 &   108.47(15) &   -84.14(17) & 1.6(1.0) & HARPS \\
6579.49121 &   122.85(15) &   -97.65(17) & 1.9(1.0) & HARPS \\
6908.52162 &   129.63(15) &  -103.69(17) & 2.4(1.0) & HARPS \\
6908.60189 &   127.83(15) &  -102.24(17) & 1.2(1.0) & HARPS \\
6909.52328 &   -94.03(15) &   107.85(17) & 2.1(1.0) & HARPS \\
6909.58819 &  -102.62(15) &   116.23(17) & 3.0(1.0) & HARPS \\
6910.53338 &    76.02(15) &   -52.86(17) & 1.4(1.0) & HARPS \\
6910.65710 &   104.42(15) &   -80.37(17) & 1.2(1.0) & HARPS \\
\hline      
\end{tabular}
\end{table}

\begin{table}
\begin{center}
\caption{Observed magnitudes of the system IO~Aqr}
\label{magnitud}
\begin{tabular}{lccccc}
\hline \hline
Band & &Ref. &\multicolumn{3}{c}{IO~Aqr}  \\
   & total(err.)&  & primary & secondary & third body \\
\hline
B               &9.315(31)&3 &10.145&10.001& 15.92$^a$\\
V               &8.834(22)&1&9.685 &$9.506$& 14.63$^a$\\
R              &8.540(45)&4,5& 9.403 & 9.209 &13.81$^a$\\
I               &8.254(19)&1& 9.128& $8.919 $ & 13.14$^b$ \\  
J$_{\rm 2MASS}$ &7.830(29)&2 &8.726& 8.494& 12.13$^b$\\
K$_{\rm 2MASS}$ &7.555(33)&2 & 8.474&8.222& 11.36$^b$\\
\hline
\end{tabular}
\\
\end{center}
{\small Source: 1 --- this work (ASAS), 2 --- \cite{cut03}, 3 --- \cite{hog00} (Tycho-2), 4 --- \cite{pic10}, 5 --- \cite{zac04} (NOMAD)\\
$^a$ estimated from spectra \\
$^b$ assuming spectral type K6 V}
\end{table}

\begin{figure}
\mbox{\includegraphics[width=0.49\textwidth]{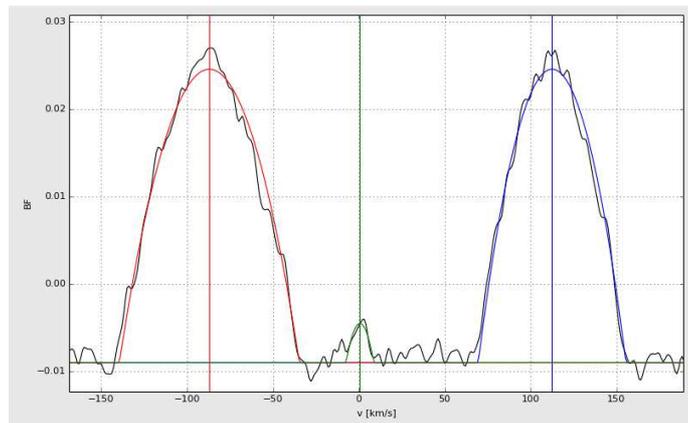}}
\caption{Broadening function power diagram of IO~Aqr's HARPS spectrum
taken 2010 September 29 (JD$=2455468.5$) during full astronomical night with
the moon well below horizon. The low-temperature template
(T$_{\rm eff}=3950$ K) was used to highlight the signal from the third body in the
spectrum - the small bump in the middle. The paraboloid curves are rotationally broadened fits to the BF of the primary (right) and the secondary (left), and the continuous vertical lines are the resulting radial velocities of the three stars.
\label{3body}}
\end{figure}

We used RaVeSpAn \citep{pil12}  to measure the  radial velocities of both
stars in the IO~Aqr binary system using the broadening function (BF)
formalism \citep{ruc92,ruc99}. We used templates from the library of synthetic
LTE spectra by \cite{col05}. We chose templates to match the
estimated effective temperature and gravity of the stars in the binary. The abundance was assumed to be solar.  
We made preliminary radial velocity fits separately to the HARPS and CORALIE data. 
There are very small differences in the systemic velocities of both components, but as they are much smaller
than dispersion of residuals, we neglected them. 
The resulting rms of the CORALIE radial velocities is about 30\% larger
than the rms of HARPS data. The line profiles of both stars are considerably
broadened by the rotation of the stars. The projected equatorial rotation
velocities of the primary and secondary components are
$v_1\sin{i}=42.6\pm1.3$\,km\,s$^{-1}$ and
$v_2\sin{i}=52.3\pm1.1$\,km\,$^{-1}$, respectively. The expected equatorial
velocities for assuming synchronous rotation are 46.9\,km\,s$^{-1}$ and 53.6\,km\,s$^{-1}$ for the primary and
the secondary, respectively, so it appears that rotation rates of both stars are very close to
synchronous rotation. The line intensity ratio at 5500 {\rm \AA} is 1.23. 

 During this analysis, we noticed an additional faint feature in the BF near
the radial velocity value 2 km s$^{-1}$ in one of the spectra -- see
Fig.~\ref{3body}. We searched for this feature in the other spectra and
confirmed its existence in every spectrum. The radial velocity of this feature
slightly changes with a spread of about 2 km s$^{-1}$. We cross-correlated
our spectra with different templates and found that the maximum strength of
the feature occurs for an assumed template temperature of $\sim4000$ K (assuming
solar metallicity). The feature contributes about 1\% of the total light at
$\sim6000 $ {\rm \AA}. Our radial velocity measurements for all three stars
are given in Table~\ref{tab_rv}. 

The detection of period variations (Sect.~\ref{per_var}) and third light in the
spectra led us to the hypothesis that there is a third body in the system,
probably a low-mass main-sequence star. This companion star
probably is responsible for the additional signal in the spectrum and the period changes
due to the light-time effect, which result from its orbital motion around the main
binary.

\subsection{Interstellar extinction\label{extinc}}
 The SIMBAD/VizieR database quotes a few different spectral type estimates for
IO~Aqr: F5 V, F5 III, F6 V, and G0. This range is too broad to be useful for
finding intrinsic average colours of the system. Instead, we used extinction
maps \citep{sch98} with recalibration by \cite{sch11} to determine the reddening in
direction of IO~Aqr. The total foreground reddening in this direction is
E(B$-$V$)=0.070 \pm 0.001$\,mag. The reddening to IO~Aqr is a fraction of this total.
To estimate the reddening to IO~Aqr, we assumed a simple axisymmetric model 
of an exponential disc for the distribution of dust within the Milky Way 
with scale and height lengths taken from \cite{dri01} and assumed a distance to IO~Aqr of $D=0.25$ kpc
(see Sect.~\ref{sect_distance}). For Galactic coordinates of IO~Aqr of $(l, b) = (47^\circ\!\!.1,
 -23^\circ\!\!.6)$ and assuming a solar distance to the Milky Way
centre of $R_0=8.3$ kpc \citep{gil09}, we obtain E(B$-$V)$_{IO~Aqr}=0.034$\,mag. 
Details of our procedure are given in \cite{such15}. \cite{maxted14}
found the scatter of the reddening to individual stars around the value
obtained from these reddening maps to be of 0.034\,mag.

Additionally, we used a calibration between the equivalent width of the interstellar
absorption sodium line NaI D1 and reddening by \cite{mun97}. We averaged our
measurements over several spectra of IO~Aqr obtained in orbital quadratures
between 2009 and 2014. The interstellar sodium line D1 has one sharp component
of constant radial velocity of $-10.55$ km s$^{-1}$ The mean equivalent width
of this line is 0.082~\AA, which~results in E(B$-$V)=0.019\,mag. The scatter of the
E(B$-$V) values for individual stars around the calibration of Munari et al.
for low reddening values is approximately 0.04\,mag. Both estimates point
toward  a small extinction in direction of IO~Aqr. Finally, we assumed
the extinction to be an average of both estimates: E(B$-$V)$=0.027 \pm 0.020$\,mag.

\subsection{Third light characterisation}
 We estimated the colour of the third light present in the spectrum using two
spectral windows: a blue one ranging from 4000 \AA~to 5000 \AA~and a red one
ranging from 6000 \AA~to 6800 \AA. These spectral windows approximately correspond to the Johnson B and Johnson-Cousins R photometric bands.
Each window was used to calculate the integrated BF profile strength of every
component visible in the spectra: $W_1$, $W_2$, $W_3$, where index 1 is for
the primary, etc. The profiles were calculated using a template with $T_{\rm
eff}=6350$ K, $\log{g}=3.00,$ and solar metallicity. We defined the third light
as $l_3 = W_3/(W_1+W_2+W_3)$.  Averaging over several spectra, we obtained
$l_3$(B)=0.0023 and $l_3$(R)=0.0078. From the observed magnitudes of IO
Aqr in the B and R bands, we calculated the magnitudes of the third light
in these bands (see Table~\ref{magnitud}). They correspond to an observed
colour (B$-$R$)=2.11$\,mag and de-reddened colour (B$-$R$)_0=2.05$.
Using Table 5 of
\cite{pec13}\footnote{\texttt{http://www.pas.rochester.edu/$\sim$emamajek/EEM\_dwarf\\\_UBVIJHK\_colours\_Teff.txt}}
, we found that this colour corresponds to a main-sequence star of spectral type
K6\,V. Assuming a distance of 255\,pc (Sect.~\ref{sect_distance}) and a reddening of  ${\rm E}({\rm B}-{\rm V}) = 0.027$
(Sect.~\ref{extinc}) to IO~Aqr, we calculated the R absolute magnitude
of the companion to be M$_R=6.87$\,mag, which corresponds to a spectral type between
K5\,V and K6\,V. Adding our estimate of the effective temperature given at the end
of Sect.~\ref{rad_vel}, we conclude that the companion has a
spectral type K6
$\pm$ 1 subclass. Table~\ref{magnitud} gives the magnitude of the companion in
different photometric bands assuming colours from \cite{pec13}. The expected
contribution of the third light in K band is $\sim$3\%.

 We found that the derived (B$-$R) colour depends only slightly on
the chosen template, being on average redder for  templates with lower effective
temperature. Assuming the same distance for IO~Aqr and the third light source,
it is unlikely that the source is of spectral type later than M0\,V because it
would then be too weak to produce a significant BF signal. On the other hand, a
star with spectral type earlier than K4\,V is also unlikely because it would
produce a much stronger BF signal that would match a hotter template ($\sim$ 4600 K). 

\subsection{Analysis of combined light and radial velocity curves \label{wd}}
 To perform this analysis we used version 2013 of the 
Wilson-Devinney program
(hereafter WD)
\citep{wil71,wil79,wil90,van07}\footnote{\texttt{ftp://ftp.astro.ufl.edu/pub/wilson/lcdc2013/}}.
We cleaned the WASP light curve with 3-$\sigma$ clipping and
repeated this cleaning
three times. Then we used every third WASP observation to form the final light
curve used in the WD analysis. This yielded light curves containing 4730
points in WASP band, 650 points in V band, 1174 points in I band, and 87 points
in H$_{\rm p}$ band. 

 We set $T_1= 6475$\,K (see Sect.~\ref{temp}) and [Fe/H]$=0$. The grid size
was set to $N=40,$ and standard albedo and gravity brightening for convective
stellar atmospheres were chosen. The stellar atmosphere option was used,
radial velocity tidal corrections were automatically applied, and no flux-level-dependent weighting was used. We assumed a circular orbit and synchronous
rotation for the two components. Logarithmic limb-darkening law were used
(Klinglesmith \& Sobieski 1970).  The starting point for the parameters of the
binary system were based on the solutions from the JKTEBOP analysis augmented 
by dynamical parameters from the preliminary solution obtained with RaVeSpAn, but we also
accounted for the influence of the third body on the orbit, as we describe below.

\begin{table}
\begin{centering}
\caption{Results of the final WD analysis including the third body.}
\label{tab_par_orb}
\begin{tabular}{lcccc}
\hline \hline
Parameter & Primary & Secondary \\
\hline
$P$ (d) & \multicolumn{2}{c}{2.3681275(33)}  \\
HJD$_0$ (d) & \multicolumn{2}{c}{2454046.0586(36)}  \\
$a (R_\odot)$ & \multicolumn{2}{c}{$11.048(7)$}   \\
$q$ & \multicolumn{2}{c}{1.0545(14)}    \\
$ i$ (deg) & \multicolumn{2}{c}{81.56(6)}  \\
$\Omega$ &6.124(32)&5.681(31)\\
$r_{\rm mean}$ &0.1983(13)&0.2257(15)\\
$\gamma$ (km~s$^{-1}$) & \multicolumn{2}{c}{8.45(11)}  \\
$T_{\rm eff}$ (K) &6475$^{a}$&6336(6) \\
$K$ (km~s$^{-1}$) & 119.80(10) & 113.61(11) \\
$L_2/L_1$(Hipparcos) &\multicolumn{2}{c}{1.165 (15)}\\
$L_2/L_1$(WASP) &\multicolumn{2}{c}{1.176 (14)}\\
$L_2/L_1$(ASAS-V)&\multicolumn{2}{c}{1.178 (14)}\\
$L_2/L_1$(ASAS-I) &\multicolumn{2}{c}{1.213 (13)}\\
$l_3$(Hipparcos) &\multicolumn{2}{c}{0.0048$^{a}$}&\\
$l_3$(WASP) &\multicolumn{2}{c}{0.0048$^{a}$}&\\
$l_3$(ASAS-V)&\multicolumn{2}{c}{0.0048$^{a}$}& \\
$l_3$(ASAS-I) &\multicolumn{2}{c}{0.0111$^{a}$}&\\
$e_3$ &\multicolumn{2}{c}{0.775(76)}\\
$\omega_3$ (deg) &\multicolumn{2}{c}{227(16)}\\
$a_3 \sin{i_3} (R_\odot)$ &\multicolumn{2}{c}{6354(30)}\\
T$_{03}$ (d)&\multicolumn{2}{c}{2449611(584)} \\
$P_3$ (d)&\multicolumn{2}{c}{30000$^{a}$}\\
$q_3=$M$_3\sin{i_3}/($M$_1+$M$_2)$ &\multicolumn{2}{c}{0.186} \\ 
RV (HARPS) $rms$ (m~s$^{-1}$) & 169 & 179  \\
RV (CORALIE) $rms$ (m~s$^{-1}$) & 209& 254 \\
Hipparcos $rms$ (mmag) &\multicolumn{2}{c}{14.6}\\
WASP $rms$ (mmag) &\multicolumn{2}{c}{7.1}\\
ASAS-V $rms$ (mmag) &\multicolumn{2}{c}{13.4}\\
ASAS-I $rms$ (mmag) &\multicolumn{2}{c}{8.4} \\
$\chi^2/DOF$  & \multicolumn{2}{c}{1.02}  \\
\hline
\end{tabular}
\end{centering}
\\$^a$ Fixed value
\end{table}

\subsubsection{Effective temperature estimates\label{temp}}
To estimate the effective temperatures of the eclipsing components, we used a
number of ($V\!-\!K$) and ($V\!-\!I$) colour - temperature calibrations
\citep{ram05,gon09,ben98,cas10,mas06,hou00}. The intrinsic colours of the two
components were calculated from observed magnitudes, estimated reddening, flux
ratios derived from the WD model, and from the third light contribution. As a source for K-band photometry in 2MASS 
\citep{cut03} (see Table~\ref{magnitud}), we used appropriate colour transformations
for each calibration. Because flux ratios from the WD model depend on the input temperatures, we iterated the estimation of
temperatures a few times. The resulting temperatures are averages from all used
calibrations. For the primary and secondary they are $6475 \pm 85$ K and
$6331 \pm 69$ K, respectively, with errors calculated as the
standard deviation.
When the reddening and zero-point uncertainties of the
photometry are added, the total errors of the temperature 
determination are 138\,K and 125\,K for the primary and the secondary, respectively.   

\subsubsection{Analysis including the third body  \label{3bod}}   
 We assumed that the changes in orbital period are caused entirely by the
light-time effect~\citep{wol22,irw59,van07}. Changes in the orbital period
of the inner binary due to gravitational interaction between the three bodies
are expected to be at least two orders of magnitude smaller than the variation
seen in Fig.~\ref{fig_ephem} \citep[][their Eq. 10]{rap13}. A
preliminary
analysis of our radial velocities showed that the systemic velocity is
$\gamma\approx10$ km s$^{-1}$. \cite{dim04} reported a much lower systemic
velocity of $-2$ km s$^{-1}$ from their observations taken in Rozen
Observatory with the 2.0 m RCC telescope, which is equipped with a Coud\'{e} spectrograph.
Their spectra were taken in 2001, so there is a difference of about nine years
to our spectra. We initially interpreted the change of the systemic velocity as a sign
of interaction between the binary and the third body. However, we quickly
realized that the magnitude of the systemic velocity change ($\sim 12$ km s$^{-1}$)
is much too large: if this were real, we would expect much larger O$-$C residuals from the
linear ephemeris given in Sect.~\ref{per_var}.  
We concluded that there is a large systematic zero-point offset in velocities
derived from our spectra and the Rozen spectra, and because their radial
velocities are on average of lower accuracy than ours, we dropped them from
our analysis.

In the analysis we adjusted the
following parameters describing the inner binary of IO~Aqr: the binary orbital period $P$, the zero-epoch of the
primary minimum HJD$_0$, the systemic velocity $\gamma$, the temperature of the
secondary $T_2$, the orbital inclination $i$, the two surface potentials
$\Omega_1$ and $\Omega_2$, the semi-major axis $a$, the mass ratio $q,$ and the
primary star luminosity $L_1$ in each photometric band.

The O$-$C diagram shows that our eclipse timing and radial
velocity data probably cover only a small fraction of the orbital
period of the third body, $P_3$. In this case, the third body orbital parameters are not
uniquely determined. A striking feature of our radial velocities is the lack of significant
changes in the systemic velocity of the inner binary or the mean radial
velocity of the third body. Moreover, the shape of the O$-$C diagram suggests a
highly eccentric and long-period third body orbit. 
It was also clear from our trial solutions that the orbital
period $P_3$ and the eccentricity $e_3$ strongly correlate, as
is the case for the longitude of
periastron $\omega_3$ and the time of the upper spectroscopic conjunction
T$_{03}$. For this reason, we decided to find solutions within a grid of
possible periods $P_3$. The additional fitted parameters were those describing
the third body orbit: eccentricity $e_3$, longitude of the periastron
$\omega_3$, semi-major axis $a_3$ , and the moment of the upper spectroscopic
conjunction T$_{03}$.  We also assumed that the outer orbit is seen edge-on
($i_3=90$ deg).

\begin{figure}
\mbox{\includegraphics[width=0.49\textwidth]{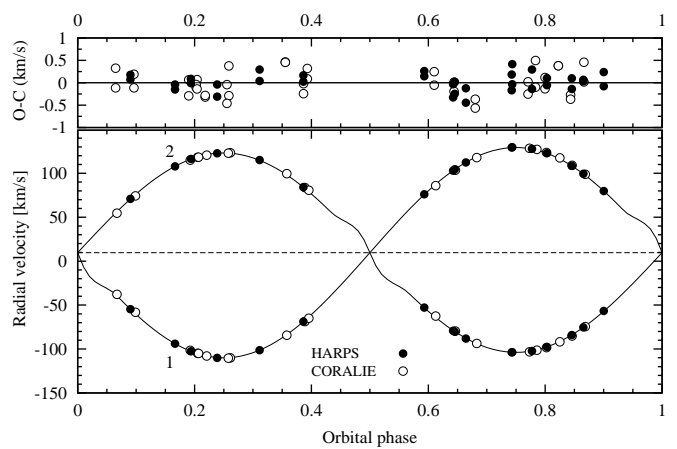}}
\caption{Radial velocity solution of IO~Aqr.  
The lower panel shows the model predictions with continous lines 
from the WD code calculated for a constant
orbital period of P=2.368137 d. Overplotted are radial velocities for 
the primary (denoted 1) and the secondary (denoted 2). 
The dotted line corresponds to the systemic velocity of 9.66 km
s$^{-1}$ corresponding to a mean epoch of CORALIE and HARPS spectra. 
The upper panel shows residuals from the model fit. 
\label{veloc}}
\end{figure}

We decided to divide the third body
parameters into three groups -- \{$e_3$, T$_{03}$\}, \{$\omega_3$\}, and
\{$a_3$, T$_{03}$\}.  We iteratively fitted these three parameter groups with
the parameters of the inner binary ($P$, HJD$_0$, $\gamma$, $T_2$, $i$,
$\Omega_1$, $\Omega_2$, $a$, $q,$ and $L_1$) until we achieved no improvement
in $\chi^2$ for a given value of $P_3$. We repeated this procedure for many
orbital periods $P_3$ within a range of from 8000 to 50000 days. 
We could not find any acceptable
third body solutions for periods shorter than about 25000 days. The reason is that the radial velocity difference between the inner binary systemic velocity and the radial velocity of the  third body ($\Delta V_{\rm r}\approx 8$ km s$^{-1}$) causes a tension with
the shape of the O$-$C curve. The tension is only relaxed for $P_3 > 25000$ days.
Periods longer than 50000 d require a very high eccentricity of the outer orbit
($e_3 > 0.85$). In all cases the inferred minimum mass of the third body is
about 0.6 $M_\odot$. The specific solution for $P_3=30000$ days is reported in
Table~\ref{tab_par_orb}. Figure~\ref{veloc} presents the fit to the radial velocity data given in
Table~\ref{tab_rv} assuming a constant orbital period during the time span of our spectroscopic
data.  

\begin{figure*}
\mbox{\includegraphics[width=0.49\textwidth]{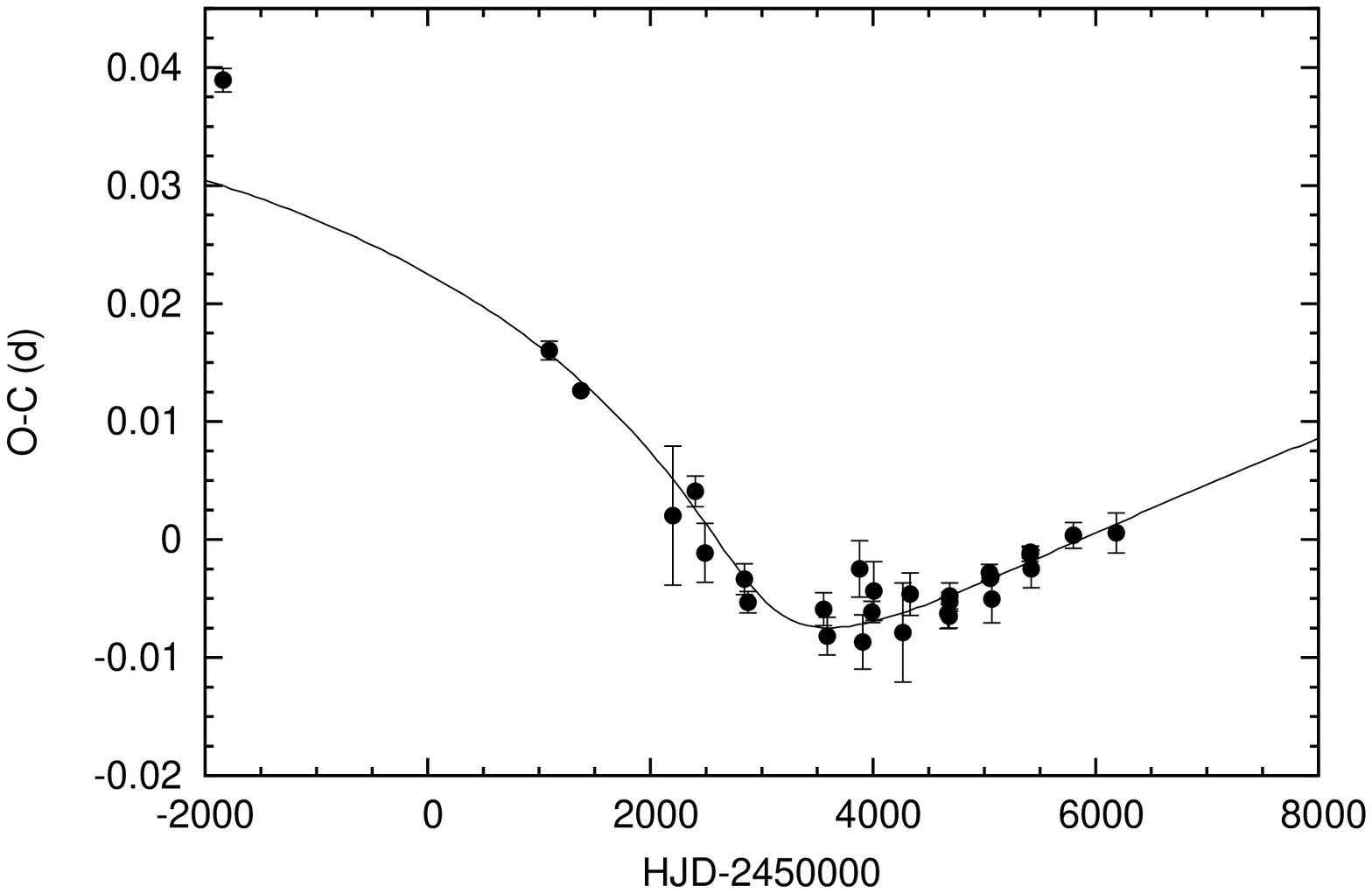}} \hfill 
\mbox{\includegraphics[width=0.49\textwidth]{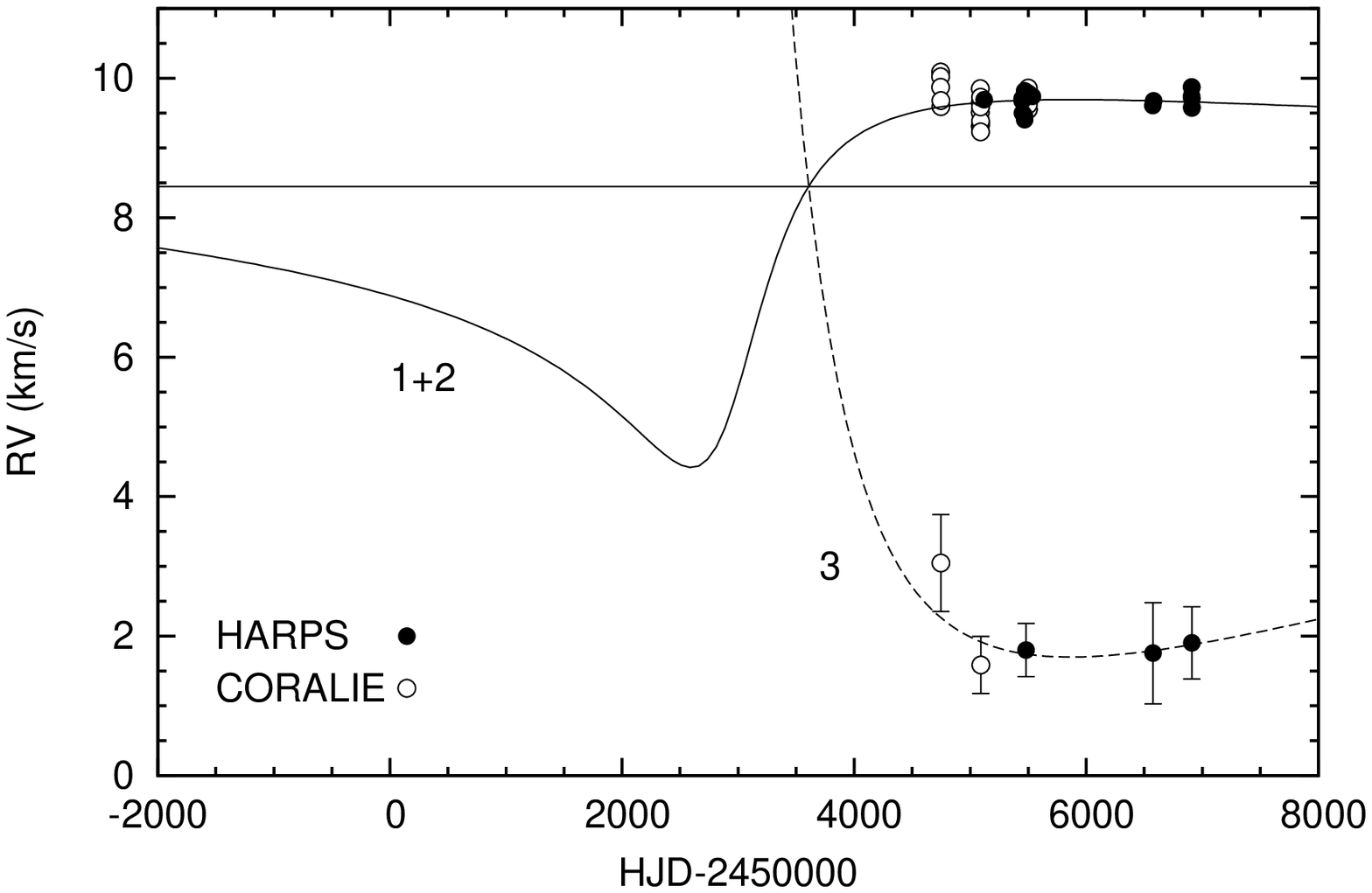}}
\caption{ {\it Left panel}: O$-$C residuals of the observed times of primary minimum of IO~Aqr
assuming the linear ephemeris from Table~\ref{tab_par_orb} and predicted O$-$C
caused by the light-time effect from our third body solution -- continous line.
{\it Right panel}: predicted radial velocity changes of the binary mass centre
(continuous line denoted 1+2) and the third body (dashed line denoted 3). 
The horizontal line corresponds to the heliocentric frame systemic velocity of the whole triple system. 
The individual binary mass centre velocities from HARPS are marked by filled circles and CORALIE velocities
by open circles. The velocities of the third body were binned into seasonal means. 
\label{3orbit}}
\end{figure*}

Figure~\ref{3orbit} presents the fit to the O$-$C residuals of observed
minima times from Table~\ref{tab_tmin} and the expected radial velocity
changes of the inner binary mass centre of IO\ Aqr and the third body. To calculate
the binary centre-of-mass velocities $V_{1+2}$ , we used radial velocities from Table~\ref{tab_rv}
and the equation
\begin{equation}
V_{1+2} = \frac{V_1 + qV_2 }{1+q}
,\end{equation}
where $V_{1,2}$ denotes radial velocities for the primary and the secondary, respectively,
and $q$ is the mass ratio. The asymmetric minimum in the O$-$C diagram coincides with the
periastron passage of the third body near JD~2452600. The O$-$C diagram shows that
our final solution for the third body orbit does not fit the
Hipparcos data. Forcing a good fit for this one outlier results
in stronger binary centre-of-mass radial velocity changes and produces significant residua in the radial velocities. 
A possible explanation is that we underestimated the error on Hipparcos minimum timing. It is also clear that we only cover a relatively small part of the whole third body orbit, and much room for improvement probably still exists.

\subsection{Physical parameters}
 The physical parameters of the components of IO~Aqr are presented in
Table~\ref{par_fi}.  Our spectral type estimate is based on the derived
effective temperatures and the calibration by \cite{pec13}; it is consistent
with the spectral classification given by \cite{hou99}. The mean radii given
are the radius of a sphere with the same volume as the star. The distortions
from a perfect sphere defined as $(r_{\rm point} - r_{\rm pole})/ r_{\rm
mean}$ are 2.3\% and 3.2\% for the primary and the secondary star,
respectively. Errors on luminosities were calculated from uncertainties of the distance estimate and the 
reddening and bolometric corrections used. The bolometric corrections reported in Table~\ref{par_fi} were taken
from \cite{cas10}. The fractional precision of derived masses is $0.25\%$ and $0.24\%$ for the primary and the secondary, respectively.

\begin{figure}
\mbox{\includegraphics[width=0.49\textwidth]{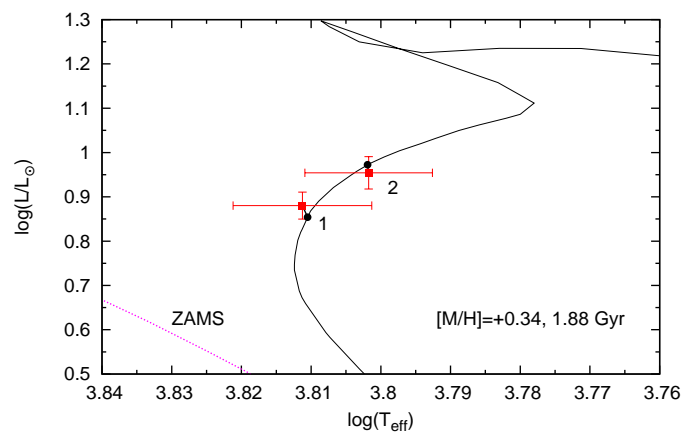}}
\caption{ H-R diagram with the primary (1) and the secondary (2) components of IO~Aqr marked. Overplotted is the PARSEC isochrone that provides the best match to $M$, $L,$ and $T_{\rm eff}$ of the two stars, corresponding to an age of 1.88 Gyr. Dots on the isochrone correspond to the position of the best fit. The zero-age main-sequence (ZAMS) is also shown.  
\label{hr}}
\end{figure}

\cite{dim04} reported two possible solutions for absolute
parameters of IO~Aqr denoted as A and B. It becomes
evident that our solution lies somewhere between these solutions. However,
we were able to significantly refine the parameters of the system. The radii of the two
stars are too large for their mass or for their temperature if we were to assume
that they are main-sequence stars, as was also pointed out by
\cite{dim04}. We compared our results with stellar isochrones from the Padova and Trieste Stellar Evolutionary Code
\citep[PARSEC;][]{bre12}\footnote{\texttt{http://stev.oapd.inaf.it/cgi-bin/cmd}}. We fitted masses, luminosities, and temperatures of the components. As we have no confident metallicity estimate for this star (the
SIMBAD database quotes some estimates made under the assumption that IO~Aqr is a single star), we also fitted this parameter. The position of the stars in the Hertzsprung-Russell (H-R) diagram and their measured masses are only matched using isochrones for metallicities ${\rm [M/H]} \approx +0.3$ - see Fig.~\ref{hr}. The estimated age of the system is   $1.88 \pm 0.10$ Gyr. We confirm the finding of \cite{dim04} that the two stars are still before the turn-off point in their evolution. 

Our absolute parameters do not depend on the orbital solution
for the third body. The third body only affects the timing of the eclipses
and, to a lesser extent, the estimates of the distance and effective
temperatures. Our solutions without the third body
(Sect.~\ref{sect_wasp_lcfit}) yield model parameters that are
consistent with those reported in Table~\ref{tab_par_orb}. The high precision
of the derived physical parameters makes IO~Aqr another eclipsing binary star on a
list of systems that have a precision of the absolute dimension determination better
than 3\% \citep[e.g.][]{tor10}. 

\begin{table}
\centering
\caption{Physical parameters of IO~Aqr components}
\label{par_fi}
\begin{tabular}{lccc}
\hline \hline
Parameter& Primary& Secondary & Tertiary \\
\hline
Spectral type & F5 V-IV &  F6 V-IV & $\sim$ K6 V\\
$M$ ($M_{\sun}$) & 1.569(4)& 1.655(4) & 0.6(1)\\
$R$ ($R_{\sun}$) & 2.191(15)& 2.493(17)& ---\\
$\log{g}$ (cgs) & 3.952(6) & 3.863(6) & ---\\
$T_{\rm eff}$ (K) & 6475(138)& 6336(125)& $\sim 4000$\\
$L$ ($L_{\sun}$) & 7.59(55)& 9.00(65) &  ---\\
$M_V$ (mag)& $2.57(8)$ & $2.39(8)$ & $\sim 7.7$\\
BC$_V$ (mag) & $-0.021$ & $-0.030$& ---\\\hline
Distance (pc) & \multicolumn{3}{c}{255(9)} \\
Distance modulus (mag) & \multicolumn{3}{c}{7.030(72)}\\
Parallax (mas) & \multicolumn{3}{c}{3.93(13)} \\
$E(B\!-\!V)$ (mag) & \multicolumn{3}{c}{0.027(20)}\\
\hline
\end{tabular}
\end{table}

\section{Distance and space velocity} \label{sect_distance}
The distance to the system was derived using two calibrations of visual surface
brightness versus $(V\!-\!K)$ colour relations \citep{ker04,ben05} based on
 measurements of the interferometric stellar angular diameters. The 2MASS $K$-band
magnitude was tranformed onto the Johnson photometric system using 
transformation equations from \cite{bes88} and
\cite{car01}\footnote{\texttt{http://www.astro.caltech.edu/$\sim$jmc/2mass/v3/\\transformations/}}.
The magnitudes were de-reddened with $E(B-V)=0.027$\,mag. To derive individual
K-band magnitudes of the components, we used the K-band light ratio extrapolated
from the WD model $l_{21}(K)=1.262$ and the third light contribution
$l_3(K)=0.030$. The resulting de-reddened Johnson V-band and K-band magnitudes for the two
stars are V$_1=9.601$\,mag, V$_2=9.422$\,mag, and K$_1=8.505$\,mag,
K$_2=8.252$\,mag. The distance to IO~Aqr derived from the SBC
relations
reported by \cite{ker04} and \cite{ben05}
are 257.2 pc and 252.2 pc, respectively. The
resulting average distance is  $255 $ pc, corresponding to a distance modulus
$m-M=7.030$\,mag.  

The systematic errors contribute as follows: the uncertainty of the empirical
calibrations of the surface brightness (0.040\,mag error in distance modulus),
the uncertainty of the extinction law (0.005\,mag), the metallicity dependence of
the surface brightness relations (0.004\,mag), and the zero-point errors of V- and
K-band photometry (0.030\,mag in total). When we combine the contributions in quadrature,
we have a total systematic error of 0.051\,mag. The total statistical error comes
from the uncertainty in the semi-major axis (0.005\,mag), the
uncertainty of the sum of
fractional radii $r_1+r_2$ (0.010\,mag), the uncertainty of the third light
(0.022\,mag), the errors of V- and K-band\,magnitudes (0.007 and 0.033\,mag,
respectively), the reddening uncertainty (0.015\,mag), the error of the mean of the
two calibrations (0.021\,mag), and from separating the magnitudes (0.002\,mag).
When we combine these uncertainties in quadrature, we have a total statistical error of
0.049\,mag. When the errors in distance modulus are translated
into errors in parsecs, the
final result is $255\pm 6$ (stat.) $\pm$ 6 (syst.) pc. The distance
corresponds to a parallax of $3.93 \pm 0.13$\,mas (total error). 

The important consistency check of model parameters we derived (especially
their radii and luminosity ratio in different bands) is the distance to each
of the system components returned by our procedure. In our case,
the distance moduli of the two stars agree excellently well:
they are different by less than 0.001\,mag for each of the surface brightness calibration used. 

The parallax to the binary was determined by the Hipparcos satellite mission.
The first published full data reduction \citep{per97} gave $5.42 \pm
1.46$\,mas. Subsequently, a new data reduction \citep{vLe07a,vLe07b} reported
a distinctly smaller parallax of $3.71 \pm 0.94$\,mas. \cite{dim04} used an
eclipsing binary bolometric flux scaling method  and derived a distance to IO~Aqr
of 256 pc, corresponding to a parallax of 3.91\,mas. However, they neither
included any third light in their considerations nor gave a distance error
estimate. Both of the latter distance determinations agree very
well within the uncertainties with our derived distance.  
 
The proper motion of IO~Aqr given by \cite{vLe07b} is
$(\mu_{\alpha}\cos{\delta}, \mu_{\delta})=(-4.59\pm1.00,
-22.42\pm0.84)$\,mas\,yr$^{-1}$. That proper motion expressed in Galactic
coordinates is $(\mu_l\cos{b},\mu_b)=(-21.7 \pm 1.2, -7.4 \pm
0.4)$\,mas\,yr$^{-1}$ using the transformations given by \cite{pol13}.  The
transverse velocity in Galactic coordinates is $(-26.2\pm1.5,-8.9\pm0.5)$ km
s$^{-1}$. Heliocentric Galactic space velocity components, that
is, not corrected
for solar peculiar motion, were calculated using the equations given by
\cite{jon87}, and we obtained (u,v,w)$=(21.8\pm0.6,-14.9\pm1.6,-11.5\pm0.8)$ km
s$^{-1}$. The values and uncertainties given do not take into account possible
systematic errors caused by a long-term proper motion drift of the inner
binary induced by the third body. At the moment, we have insufficient
information to correct for this effect.  

\section{Final remarks}
With the high-precision data now available, we have been able to show that IO~Aqr is a hierarchical triple star system with a low-luminosity and low-mass companion on a wide eccentric orbit.
For the purpose of accurate SBC calibration, pure eclipsing binaries (i.e. without stellar companions) are preferred. However, when the physical properties, especially the luminosity, of a tertiary companion are well determined or/and the flux of the companion
in optical and NIR is negligible, a given eclipsing binary can be used safely. A criterion is the degree of calibration precision we wish to obtain: 1\% of precision demands that the tertiary contribution in V and K bands should be at most comparable with this limit. The companion of IO~Aqr is faint enough in the optical to be ignored, and its estimated K-band contribution is still small, so that it has a negligible effect on the calculated angular diameters of the IO~Aqr components. Using the SBC calibration
of \cite{ben05}, the predicted angular diameters are $0.081\pm0.002$\,mas and $0.092\pm0.002$\,mas for the primary and the secondary, respectively.   

We derived high-precision absolute dimensions of IO~Aqr that are among the most precise parameters of F-type stars published to date. Nonetheless, our temperature determination (precision ~2\%) can be improved by using dedicated Str\"omgren photometry or by a detailed atmospheric analysis of the separated spectra. We leave this to future work on IO~Aqr.

\begin{acknowledgements}

We extensively used the SIMBAD/Vizier database in our research. We also used the {\it uncertainties} python package. 
We would like to thank the referee D. Pourbaix for his constructive comments on the paper and the staff of the ESO La Silla observatory for their support during the observations. We also thank W. van Hamme for his valuable comments about modelling of a third body with the Wilson-Devinney code. 

We [D.G., W. G., G.P.] gratefully acknowledge financial support for this work from the BASAL Centro de Astrofisica y Tecnologias Afines (CATA) PFB-06/2007, and from the Millenium Institute of Astrophysics (MAS) of the Iniciativa Cientifica Milenio del Ministerio de Economia, Fomento y Turismo de Chile, project IC120009. A.G. acknowledges support from FONDECYT grant 3130361.

We [D.G., B.P., G.P., P.K., K.S.] gratefully acknowledge financial support for this work
from the Polish National Science Center grant MAESTRO
2012/06/A/ST9/00269, the TEAM subsidy from the
Foundation for Polish Science (FNP) and NCN grant DEC-2011/03/B/ST9/02573. 
R.I.~A. acknowledges funding from the Swiss National Science Foundation.
\end{acknowledgements}


\label{lastpage}


\begin{thebibliography}{99}
\bibitem[Andersen et al.(1988)]{and88} Andersen, J., Clausen, J. V., Gustafsson, B., et al. 1988, A\&A, 196, 128
\bibitem[Bessell \& Brett(1988)]{bes88}  Bessell M. S., Brett J. M., 1988, PASP, 100, 1134
\bibitem[Bressan et al.(2012)]{bre12} Bressan, A., Marigo, P., Girardi, L., et al., 2012, MNRAS, 427, 127
\bibitem[Carpenter(2001)]{car01} Carpenter J. M., 2001, AJ, 121, 2851 
\bibitem[Casagrande et al.(2010)]{cas10} Casagrande, L., Ramirez, I., Mel{\'e}ndez, J., Bessell, M., \& Asplund, M.  2010,  A\&A, 512, 54
\bibitem[Casagrande et al.(2014)]{cas14} Casagrande, L., Portinari, L., Glass, I. S., et al., 2014, A\&A, 439, 2060
\bibitem[Challouf et al.(2014)]{cha14} Challouf, M., Nardetto, N., Mourard, D., et al., 2014, A\&A, 570, 104
\bibitem[Claret et al.(1995)]{Claret95} Claret, A., Diaz-Cordoves, J.,Gimenez, A. 1995, A\&AS, 114, 247
\bibitem[Claret (2000)]{Claret00} Claret, A., 2000, A\&A, 363, 1081
\bibitem[Coelho et al.(2005)]{col05} Coelho, P., Barbuy, B., Mel{\'e}ndez, J., Schiavon, R. P., \& Castilho, B. V.  2005, A\&A, 443, 735
\bibitem[Cutri et al.(2003)]{cut03} Cutri, R. M., et al., 2003, VizieR, Online Data Catalogue, 2246, 0 
\bibitem[de Bruijne et al.(2015)]{deB15} de Bruijne, J. H. J., Rygl, K. L. J., Antoja, T., 2015, arXiv:1502.00791 
\bibitem[D\'{i}az-Cordov\'{e}s et al.(1995)]{DiazCo95} Diaz-Cordoves, J., Claret, A., Gimenez, A. 1995, A\&AS, 110, 329
\bibitem[di Benedetto(1998)]{ben98} di Benedetto, G. P.  1998, A\&A, 339, 858
\bibitem[di Benedetto(2005)]{ben05} di Benedetto, G. P.  2005, MNRAS, 357, 174
\bibitem[Dimitrov et al.(2004)]{dim04} Dimitrov, W., Kolev, D., Schwarzenberg-Czerny, A., 2004, A\&A, 417, 689
\bibitem[Drimmel \& Spergel(2001)]{dri01} Drimmel, R. \& Spergel, D. N., 2001, ApJ, 556, 181
\bibitem[Eker et al.(2014)]{eke14} Eker, Z., Bilit, S., Soydugan, F., et al., 2014, PASA, 31, 24
\bibitem[Gillessen et al.(2009)]{gil09} Gillessen, S., Eisenhauer, F., Fritz, T. K., et al., 2009, ApJ, 707, L114
\bibitem[Gonz{\'a}lez Hern{\'a}ndez \& Bonifacio(2009)]{gon09} Gonz{\'a}lez Hern{\'a}ndez, J. I., \& Bonifacio, P.  2009, A\&A, 497, 497
\bibitem[Graczyk et al.(2014)]{gra14} Graczyk, D., Pietrzy{\'n}ski, G., Thompson, I. B., et al., 2014, ApJ, 780, 59
\bibitem[H{\o}g et al.(2000)]{hog00} H{\o}g, E., Fabricius, C., Makarov, V. V., et al., 2000, A\&A, 357, 367 
\bibitem[Houdashelt et al.(2000)]{hou00} Houdashelt, M. L., Bell, R. A., \& Sweigert, A, V.  2000, AJ, 119, 1448
\bibitem[Houk \& Swift(1999)]{hou99} Houk N., \& Swift C., 1999, Michigan Catalogue of Two-dimensional Spectral Types for the HD Stars Vol. 5. Declinations $-$12 deg to +05 deg. Department of Astronomy, University of Michigan, Ann Arbor, MI, USA
\bibitem[Irwin(1959)]{irw59} Irwin, J. B. 1959, AJ, 64, 149
\bibitem[Johnson \& Soderblom(1987)]{jon87} Johnson, D. R. H., \& Soderblom, D. R., 1987, AJ, 93, 864
\bibitem[Kazarovets et al.(1999)]{kaz99} Kazarovets, A. V., Samus, N. N., Durlevich, O. V., et al., 1999, IBVS, No. 4659  
\bibitem[Kervella et al.(2004)]{ker04} Kervella, P., Th{\'e}venin, F., Di Folco, E., S{\'e}gransan, D., 2004, A\&A, 426, 297
\bibitem[Kruszewski \& Semeniuk(1999)]{kru99} Kruszewski, A., \& Semeniuk, I., 1999, AcA, 49, 561
\bibitem[Lacy(1977)]{lac77} Lacy, C. H., 1977, ApJ, 213, 458
\bibitem[Masana et al.(2006)]{mas06} Masana, E., Jordi, C., \& Ribas, I.  2006, A\&A 450, 735
\bibitem[Maxted et al.(2014)]{maxted14} Maxted, P. F. L., et al., 2014, MNRAS
437, 1681
\bibitem[Mayor et al.(2003)]{may03} Mayor M., et al., 2003, The Messenger, 114, 20
\bibitem[Munari \& Zwitter(1997)]{mun97} Munari U. \& Zwitter, T., 1997, A\&A, 318, 269
\bibitem[Pecaut \& Mamajek(2013)]{pec13} Pecaut, M. J. \& Mamajek, E. E., 2013, ApJS, 208, 9
\bibitem[Perryman et al.(1997)]{per97} Perryman, M. A. C., Lindegren, L., Kovalevsky, J., et al., 1997, A\&A, 323, L49
\bibitem[Perryman et al.(2001)]{per01} Perryman, M. A. C., de Boer, K. S., Gilmore, G., et al. 2001, A\&A, 369, 339 
\bibitem[Pickles \& Depagne(2010)]{pic10} Pickles, A. \& Depagne, E., 2010, PASP, 122, 1437 
\bibitem[Pietrzy{\'n}ski et al.(2013)]{pie13} Pietrzy{\'n}ski, G., Graczyk, D., Gieren, W., et al., 2013, Nature, 495, 76
\bibitem[Pilecki et al.(2012)]{pil12} Pilecki, B., Konorski, P., \& G{\'o}rski, M. 2012, {From Interacting Binaries to Exoplanets, IAU Symposium}, 282, 301
\bibitem[Pojma\'nski(2002)]{poj02} Pojma\'nski G., 2002, AcA, 52, 397
\bibitem[Poleski(2013)]{pol13} Poleski, R., 2013, arXiv1306.2945
\bibitem[Pollacco et al.(2006)]{pol06} Pollacco D. L., et al., 2006, PASP, 118, 1407
\bibitem[Popper \& Etzel(1981)]{pop81} Popper D. M., Etzel P. B., 1981, AJ, 86, 102
\bibitem[Ram{\'i}rez \& Mel{\'e}ndez(2005)]{ram05} Ram{\'i}rez, I., \& Mel{\'e}ndez, J.  2005, ApJ, 626, 465
\bibitem[R\'o\.zyczka et al.(2009)]{roz09} R\'o\.zyczka M., Ka\l u\.zny J., Pietrukowicz P., Pych W., Mazur B., Catel\'an M., Thompson I. B., 2009, AcA, 59, 385
\bibitem[Rappaport et al. (2013)]{rap13} Rappaport, S. et al., 2013, ApJ, 768,
33
\bibitem[Rucinski(1992)]{ruc92} Rucinski, S. M. 1992, AJ, 104, 1968
\bibitem[Rucinski(1999)]{ruc99} Rucinski, S. M. 1999, {in Precise Stellar Radial Velocities, ASP Conference Series 185, IAU Colloquium 170, ed. J. B. Hearnshaw \& C. D. Scarfe}, 82  
\bibitem[Schlafly \& Finkbeiner(2011)]{sch11} Schlafly, E. F. \& Finkbeiner, D. P., 2011, ApJ, 737, 103
\bibitem[Schlegel et al.(1998)]{sch98} Schlegel, D. J., Finkbeiner, D. P. \& Davis, M., 1998, ApJ, 500, 525
\bibitem[Southworth(2014)]{sou14} Southworth J., 2008, arXiv:1411.5517
\bibitem[Southworth et al.(2004a)]{sou04a} Southworth J., Maxted P. F. L., Smalley B., 2004a, MNRAS, 351, 1277
\bibitem[Southworth et al.(2004b)]{sou04b} Southworth J., Zucker S., Maxted P. F. L., Smalley B., 2004b, MNRAS, 355, 986
\bibitem[Stebbins(1910)]{ste10} Stebbins, J., 1910, ApJ, 32, 185
\bibitem[Stebbins(1911)]{ste11} Stebbins, J., 1911, ApJ, 34, 112 
\bibitem[Suchomska et al.(2015)]{such15} Suchomska, K., Graczyk, D., Smolec, R., et al., 2015, arXive:1505.00766
\bibitem[\protect\citeauthoryear{{Tamuz}, {Mazeh} \& {Zucker}}{{Tamuz}
  et~al.}{2005}]{2005MNRAS.356.1466T}
{Tamuz} O.,  {Mazeh} T.,    {Zucker} S.,  2005, MNRAS, 356, 1466
\bibitem[Torres et al.(2010)]{tor10} Torres, G., Andersen, J., \& Gim{\'e}nez, A., 2010, A\&AR, 18, 67
\bibitem[van Hamme(1993)]{VanHam93} van~Hamme W., 1993, AJ, 106, 2096
\bibitem[van Hamme \& Wilson(2007)]{van07} van Hamme, W., \& Wilson, R. E.  2007, ApJ, 661, 1129
\bibitem[van Leeuwen(2007a)]{vLe07a} van Leeuwen, F., 2007a, Hipparcos, the new reduction of the raw data (Dordrecht: Springer)
\bibitem[van Leeuwen(2007b)]{vLe07b} van Leeuwen, F., 2007b, A\&A, 474, 653
\bibitem[\protect\citeauthoryear{{Wilson} et~al.,}{{Wilson} et~al.}{2008}]{2008ApJ...675L.113W} {Wilson} D.~M.,  et~al., 2008, ApJL, 675, L113
\bibitem[Wilson \& Devinney(1971)]{wil71} Wilson, R. E., \& Devinney, E. J.  1971, ApJ, 166, 605
\bibitem[Wilson(1979)]{wil79} Wilson, R. E.  1979, ApJ, 234, 1054
\bibitem[Wilson(1990)]{wil90} Wilson, R. E.  1990, ApJ, 356, 613
\bibitem[Woltjer(1922)]{wol22} Woltjer, J. 1922, B.A.N., 1, 93
\bibitem[Zacharias et al. (2004)]{zac04} Zacharias, N., Monet, D. G., Levine, S. E., et al., 2004, BAAS, 36 ,1418
\end{thebibliography}
\end{document}